\documentclass[%
 reprint,
 amsmath,amssymb,
 aps,
]{revtex4-2}

\usepackage[T1]{fontenc}
\usepackage{graphicx}
\usepackage{float}
\usepackage{dcolumn}
\usepackage{bm}
\usepackage{multirow}
\usepackage{amsmath, amsthm, amsfonts,mathptmx}
\usepackage{algpseudocode}
\usepackage{fullpage}
\usepackage{times}
\usepackage{fancyhdr,amssymb}
\usepackage[ruled,longend,linesnumbered]{algorithm2e}
\usepackage[colorlinks,linkcolor=blue,citecolor=blue,urlcolor=blue]{hyperref}
\usepackage{enumerate,makecell,xcolor,xspace,stmaryrd}
\usepackage{array}
\usepackage[caption=false,font=normalsize,labelfont=sf,textfont=sf]{subfig}
\usepackage{textcomp}
\usepackage{stfloats}
\usepackage{url}
\usepackage{verbatim}
\def\BibTeX{{\rm B\kern-.05em{\sc i\kern-.025em b}\kern-.08em
    T\kern-.1667em\lower.7ex\hbox{E}\kern-.125emX}}
\usepackage{balance}
\definecolor{addn}{RGB}{255, 0, 0}

\newcommand{\ie}{\textit{i.e.}\xspace}

\begin{document}

\preprint{APS/123-QED}

\title{Quantum ($t$,$n$) Threshold Multi-Secret Sharing based on Cluster States}

\author{Rui-Hai Ma$^{1}$, Hui-Nan Chen$^{1}$}
\author{Bin-Bin Cai$^{1,2}$}
\email{Corresponding author: cbb@fjnu.edu.cn}
\author{Song Lin$^{1}$}
\email{Corresponding author: lins95@gmail.com}
\author{Xiao-Chen Zhang$^{1}$}

\address{$^1$ College of Computer and Cyber Security, Fujian Normal University, Fuzhou 350117, China\\
$^{2}$ Digital Fujian Internet-of-Things Laboratory of Environmental Monitoring, Fujian Normal University, Fuzhou 350117, China}

\begin{abstract}
Quantum secret sharing is an encryption technique based on quantum mechanics, which utilizes  uncertainty principle to achieve security in transmission. Most protocols focus on the study of quantum ($n,n$) or ($t,n$) threshold single secret sharing. In this paper, the first quantum ($t,n$) threshold multi-secret sharing protocol based on Lagrangian interpolation and cluster states is proposed, which requires only $t$ instead of $n$ participants to reconstruct multiple quantum secrets. The protocol exploits the security properties of the cluster state to transmit shared information in two parts, quantum and classical, where the shares remain private after reconstructing quantum secrets. Meanwhile, extending the new measurement basis in cluster states enables participants to transmit quantum information without preparing particles. In the presented protocol, the dealer can be offline after sending secrets. And required quantum operations are all common quantum operations, thus the protocol is practical under the current technical conditions. It is proven to be theoretically secure against external and internal attacks by analyzing the protocol under several common external attacks and internal attacks. In addition, experiments on IMB Q prove that the protocol satisfies correctness and feasibility.
\end{abstract}

\maketitle


\section{Introduction}\label{Sec1}

In view of the principles of quantum superposition, quantum entanglement and quantum interference, quantum mechanics has emerged abundant applications in cryptography \cite{bb84,zhou2004,qin2024}, machine learning \cite{harrow2009,rebentrost2018,song2024} and cryptanalysis \cite{shor1999,grover1996,simon1997}. One of the most important branch in quantum cryptography is quantum secret sharing. In 1999, Hillery et al$.$ \cite{Hillery1999} introduced secret sharing \cite{Blakley1979,Shamir} into the quantum domain by proposing an ($n,n$) threshold quantum secret sharing protocol based on Greenberger-Horne-Zeilinger (GHZ) states. Since then, a large number of threshold quantum secret sharing protocols \cite{Dehkordi2013,Qin2015,Sutradhar2021,Wang2022,Li2022,Rathi2023,Guan2023,Singh2005,Yan2022,Song2017,LiF2022,Deng2005,Xiao2004,Zhang2005,Yu2008,Keet2010,Tavakoli2015,Pinell2020,Yi2021,Bai2021,Mashhadi2022,MaS2023} have been studied and proposed.
In 2013, Dehkordi et al$.$ \cite{Dehkordi2013} proposed a (2,3) threshold quantum secret sharing scheme based on GHZ states, Lagrange interpolation and unitary operations. In 2015, Qin et al$.$ \cite{Qin2015} designed a protocol applying the phase shift operations and Lagrange interpolation to achieve a ($t,n$) threshold structure for quantum secret sharing. In 2021, Sutradhar et al$.$ \cite{Sutradhar2021} achieved high dimensional threshold quantum secret sharing based on quantum Fourier transform as well as Lagrange interpolation.
Later, Wang et al$.$ \cite{Wang2022} proposed a threshold quantum secret sharing protocol based on the quantum walk and unitary transforming operations.
In 2022, Li et al$.$ \cite{Li2022} implemented a dynamic threshold quantum secret sharing based on high dimensional Bell states. In 2023, Rathi et al$.$ \cite{Rathi2023} presented a threshold quantum secret sharing scheme based on high dimensional Bell states and single qubit unitary operations.
Recently, Guan et al$.$ \cite{Guan2023} proposed a quantum threshold secret sharing protocol based on high dimensional single quantum qubits unitary operations.

It is worth noting that all of the above schemes are single secret sharing, and in some cases it is inefficient to adopt only single secret sharing scheme. For instances, when multiple secrets needs to be transmitted. If each secret is shared in the form of single secret sharing scheme, there will be excessive information of shares need to be saved. Whereas multiple secret sharing protocols can simply utilize one set of shares (or some possible additional information) to reconstruct multiple secrets, which improves efficiency of transmission. In 1994, Blundo et al$.$ \cite{Blundo1994} firstly discussed the concept of multi-secret sharing schemes, and proposed a general theory of multi-secret sharing schemes using an information theoretic framework.
In 2000, Chien et al$.$ \cite{Chien2000} proposed a multi-secret sharing scheme based on systematic block codes. In 2004, Yang et al$.$ \cite{Yang2004} improved the scheme of Chien et al$.$ \cite{Chien2000} and presented a multi-secret sharing scheme based on Lagrange interpolation and less additional public information. In 2007, Zhao et al$.$ \cite{Zhao2007} proposed a relatively efficient verifiable multi-secret scheme based on the scheme of Yang et al$.$ \cite{Yang2004}  and combined with discrete logarithms.
In 2008, Dehkordi et al$.$ \cite{Dehkordi2008} proposed a threshold verifiable multi-secret scheme based on the scheme of Yang et al$.$ \cite{Yang2004} combined with discrete logarithms and with the RSA cryptosystem \cite{RSA}. Up until 2008, the research on multi-secret sharing \cite{Blundo1994,Chien2000,Yang2004,Zhao2007,Dehkordi2008} has been well established and mainly focuses on the field of classical information.

Currently, the existing quantum secret sharing protocols only focus on  ($t,n$) threshold \cite{Sutradhar2021,Wang2022,Li2022,Rathi2023,Guan2023,Singh2005,Yan2022,Song2017,LiF2022} or ($n,n$) threshold \cite{Deng2005,Xiao2004,Zhang2005,Yu2008,Keet2010,Tavakoli2015,Pinell2020,Yi2021,Bai2021,Mashhadi2022,MaS2023} single secret sharing. There is no studies on the combination of ($t,n$) threshold and multi-secret sharing. To achieve flexible access to multiple quantum secrets, a ($t,n$) threshold quantum multi-secret sharing  is proposed in this paper.
The proposed protocol takes advantage of quantum entanglement as well as Lagrange interpolation properties to transmit information through both quantum and classical channels, and only $t$ users ($t\leq n$) are needed to reconstruct multiple quantum secrets.
In addition, a new measurement basis is presented in this paper, which enables the participants, except for the secret dealer and reconstructor, to transmit the cluster state information only by measuring particles rather than preparing particles. The secret dealer can go offline after sending the encrypted quantum secret, and does not need to participate in other steps of the protocol. At the same time, the quantum operations required by the dealer and reconstructor are all common quantum operations, hence the protocol is practical under the existing technical conditions. Through security analyses, it is shown that the proposed protocol is secure against several common external and internal attacks.  Furthermore, the experiments on IMB Q are conducted to demonstrate the process of multi-secret sharing, proving   the correctness and feasibility of the proposed protocol.

The rest of the paper is organized as follows. In Sec$.$ \ref{Sec2}, we briefly review the cluster state and its property. Then, the process of protocol and a simple example are described in Sec$.$ \ref{Sec3}. The correctness and security are analyzed in Sec$.$ \ref{Sec4}. A simulation experiment is provided in Sec$.$ \ref{Sec5}. In the end, a conclusion is given in Sec$.$ \ref{Sec6}.

\section{Preliminary}\label{Sec2}

Cluster states \cite{Briegel2001,cluster1,Nielsen2006} are highly entangled quantum states with applications in quantum  secret sharing. Each particle of a cluster state is entangled with its neighbouring particles. The information of system state can be extracted by single particle measurements. And simultaneously the relevant quantum operations are applied to the cluster states.

\begin{figure}[h]
    \centering
    \includegraphics[width=1 in]{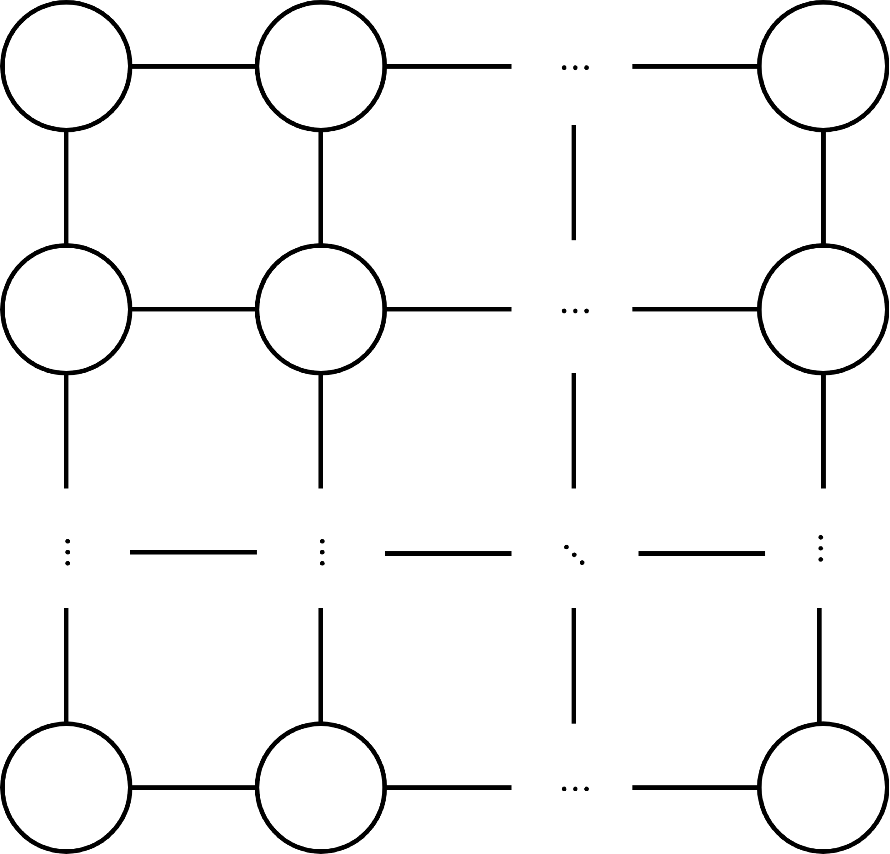}
    \caption{Graphical representation of the cluster state.}
    \label{fig1}
\end{figure}

Cluster states can be represented in the form of graphs, denoted by $G(V,{}E)$, as shown in Figure \ref{fig1}. Here, $V$ is the set of all nodes in graph $G$, and $E$ is the set of all edges. In graph $G$, each node represents a qubit in the state $|+\rangle=H|0\rangle=1/\sqrt{2}(|0\rangle+|1\rangle)$, and each edge represents a controlled-Z operation $CZ$, where operations $H$ and $CZ$ are shown in the following equations,

\begin{equation}\label{HCZ}
    \begin{aligned}
        H=\frac{1}{\sqrt{2}}\begin{bmatrix}
            1 & 1  \\
            1 & -1
        \end{bmatrix}, \\
        CZ=\begin{bmatrix}
            1 & 0 & 0 & 0 \\
            0 & 1 & 0 & 0 \\
            0 & 0 & 1 & 0 \\
            0 & 0& 0 & -1
        \end{bmatrix}.
    \end{aligned}
\end{equation}
The cluster state $|C\rangle$ has some favorable properties for secret reconstruction and satisfies the following characteristic equation \cite{Briegel2001},
\begin{equation}
    K |C \rangle = |C\rangle,
    \label{eigen_equation}
\end{equation}

\noindent where $K = X_{a} \mathop{\otimes}\limits_{b\in N(a)} Z_{b}\mathop{\otimes}\limits_{c \in (V-a-N(a))} I_c$ is the stabilizer operator for
\begin{equation}
    \begin{aligned}
        X&=\begin{bmatrix}
            0 & 1 \\
            1 & 0
        \end{bmatrix},
        Z=\begin{bmatrix}
            1 & 0 \\
            0 & -1
        \end{bmatrix},
        I=\begin{bmatrix}
            1 & 0 \\
            0 & 1
        \end{bmatrix}.\\
    \end{aligned}
    \label{IXZ}
\end{equation}
The subscript $a$ denotes an arbitrary node in the cluster state, $N(a)$ denotes the set of neighboring nodes connected to node $a$, and $(V-a-N(a))$ denotes the set of other nodes in the cluster state. In addition, $R_X(\theta)$ and $R_Z(\theta)$ are the rotation operations corresponding to $X$ and $Z$, respectively, as shown in the following equations,
\begin{equation}
    \begin{aligned}
        R_X(\theta)&=\begin{bmatrix}
            \mathrm{cos\frac{\theta}{2}} & -\mathbf{i} \mathrm{sin\frac{\theta}{2}} \\
            -\mathbf{i}{\mathrm{sin}\frac{\theta}{2}} & \mathrm{cos\frac{\theta}{2}}
        \end{bmatrix},
        R_Z(\theta)=\begin{bmatrix}
            e^{-\mathbf{i}\frac{\theta}{2}} & 0 \\
            0 & e^{\mathbf{i}\frac{\theta}{2}}
        \end{bmatrix},
    \end{aligned}
    \label{rxz}
\end{equation}
where $\theta\in [0,2\pi]$ is an arbitrary rotation angle and $\mathbf{i}=\sqrt{-1}$.

The basic state in cluster states is $|\pm\rangle$ with angle $\omega$, noted as $|\pm_\omega\rangle$, which satisfies
\begin{equation}
    \begin{aligned}
        |\pm_\omega\rangle = R_Z(\omega)|\pm\rangle = \frac{1}{\sqrt{2}}(e^{-\mathbf{i}\frac{\omega}{2}}|0\rangle)\pm
        e^{\mathbf{i}\frac{\omega}{2}}|1\rangle).
    \end{aligned}
    \label{basic particle}
\end{equation}
The new measurement basis proposed in this paper is $\{|0_{-\omega}\rangle,|1_{-\omega}\rangle\}$, where
\begin{equation}
    \begin{aligned}
        |0_{-\omega}\rangle=R_X(-\omega)|0\rangle,|1_{-\omega}\rangle=R_X(-\omega)|1\rangle.
    \end{aligned}
    \label{basis01}
\end{equation}
Measuring a particle of two-qubit cluster states in the basis $\{|0_{-\omega}\rangle,|1_{-\omega}\rangle\}$ satisfies
\begin{equation}
    \begin{aligned}
                C{Z_{1,2}}| + {\rangle _1}| + {\rangle _2} =  \frac{1}{\sqrt{2}}(|0_{-\omega}\rangle_1 |+_{\omega}\rangle_2  + |{1_{-\omega}}\rangle_1 |-_{\omega}\rangle_2).
    \end{aligned}
    \label{new_measure}
\end{equation}

The commutative formulas for $Z$, $R_Z(\varphi)$, $X$ ,$R_X(\varphi)$ and $H$ involved in this paper are
\begin{equation}\label{eqcite1}
    \begin{aligned}
        ZH = HX, XH = HZ,\\
        {R_Z}(\varphi )H = H{R_X}(\varphi ),\\
        {R_Z}(\varphi )X = X{R_Z}( - \varphi ).
    \end{aligned}
\end{equation}

\section{Protocol}\label{Sec3}

\begin{figure}[h]
    \centering
    \includegraphics[width=2 in]{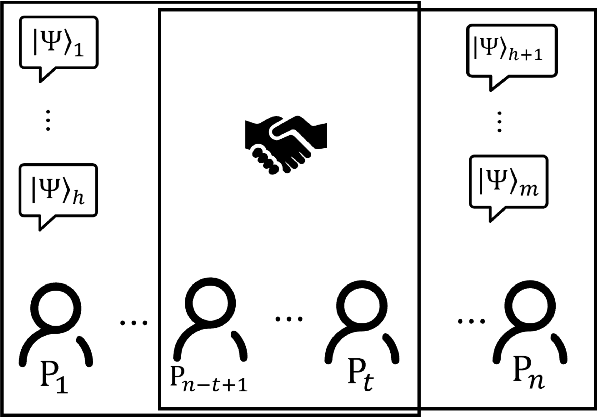}
    \caption{The macro framework of the protocol. In the figure, any $t$ rather than  $n$ users can cooperate to reconstruct multiple secrets, such as $\{|\Psi\rangle_1$, $\cdots$, $|\Psi\rangle_{h}\}$ or $\{|\Psi\rangle_{h+1}$, $\cdots$, $|\Psi\rangle_{m}\}$, where $m \geq h\geq 1$.}
    \label{fig_frame}
\end{figure}

In the presented protocol, it enables secure transmission of multi-quantum secrets from a secret dealer to a secret reconstructor. The participants consist of a Dealer and $n$ users \rm{P}$_1$, \rm{P}$_1$, ..., \rm{P}$_n$, where any $t$ users can cooperate to reconstruct the secret. The protocol is divided into two phases, secret splitting and secret reconstruction. One set of shares is able to reconstruct multiple quantum secrets. For the reader's convenience, the overall framework of the protocol is shown in Figure \ref{fig_frame}.

\subsection{Secret Splitting Phase}

\textbf{(S$_1$)} Dealer randomly chooses a number $s_{D}$ in a Galois field GF(q), where q $>n$. And Dealer constructs a polynomial
$f(x) = {a_0} + {a_1}{x^1} +  \cdots  + {a_{t - 1}}{x^{t - 1}}$ of degree $t-1$, where parameters ${a_0} = f(0) = \rm{q}$ $- s_D$, $\{ {a_0},{a_1},...,{a_{t - 1}}\}  \in GF(\rm{q})$ are coefficients. Dealer computes $f({x_i})\mod\rm{q}$, and transmits them to the corresponding user \rm{P}$_i$ via a quantum key distribution protocol \cite{bb84}, where ${x_i}$ are the public information related to the user \rm{P}$_i$ and $i \in \{ 1,2,...,n\}$. Without loss of generality, assume that users \rm{P}$_1$, \rm{P}$_1$, ..., \rm{P}$_t$ cooperate  and one of the users, such as \rm{P}$_t$ (referred to as ``secret reconstructor''), reconstructs the sequence of quantum secrets ${\{ |\psi \rangle _1},|\psi {\rangle _2}, \ldots ,|\psi {\rangle _m}\}$. Furthermore, Dealer chooses parameters ${w_j} \in \{ 1,2,...,\rm{q}- 1\} $ for each secret $|\psi\rangle_j$ and publishes them to prevent the internal attacks, where $j\in\{1,2,...,m\}$.

\textbf{(S$_2$)} The users ${\rm{P}_1},{\rm{P}_2},...,\rm{P}$$_t$ compute their parameters
\begin{equation}
    \begin{aligned}
        {c_l} = f({x_l})\mathop \prod \limits_{v = 1,v \ne l}^t \frac{{ - {x_v}}}{{{x_l} - {x_v}}}\quad \bmod \,\rm{q},
    \end{aligned}
    \label{ci}
\end{equation}
where $l = 1,2,...,t$. Then the number $s_D$ and the parameters $c_l$ satisfy
\begin{equation}
    \begin{aligned}
        ({s_D} + \mathop \sum \limits_{l = 1}^t {c_l})\quad \bmod \,\rm{q} = 0.
    \end{aligned}
\end{equation}
Dealer and users \rm{P}$_l$ calculate the rotation angles $\gamma _D^j$ and $\gamma _l^j$ based on $w_j$, $s_D$ and $c_l$,
\begin{equation}
    \begin{aligned}
        &\gamma _D^j = {w_j} \cdot \frac{{{s_D}}}{\rm{q}} \cdot 2\pi,\\
        &\gamma _l^j = {w_j} \cdot \frac{{{c_l}}}{\rm{q}} \cdot 2\pi.
    \end{aligned}
    \label{angle}
\end{equation}
The angles $\gamma _D^j$ and $\gamma _l^j$ satisfy the following equation,
\begin{equation}
    \begin{aligned}
        \gamma _D^j + \sum\limits_{l = 1}^t {\gamma _l^j}  = 2\pi r,
    \end{aligned}
    \label{sumeq2pi}
\end{equation}
where the number $r$ is an integer.

\textbf{(S$_3$)} Dealer applies the rotation operation ${R_X}(\gamma _D^j)$ to encrypt the quantum secret ${\{ |\psi \rangle _1},|\psi {\rangle _2}, \ldots ,|\psi {\rangle _m}\} $ into the encrypted quantum secrets ${\{ |\Psi \rangle _1},|\Psi {\rangle _2}, \ldots ,|\Psi {\rangle _m}\}$, \ie, $|\Psi {\rangle _j} = {R_X}(\gamma _D^j)|\psi {\rangle _j}$, where $j \in \{ 1,2,...,m\} $. Dealer sends them to the secret reconstructor \rm{P}$_t$ with decoy particles in BB84 states \cite{bb84}, \ie,  $|0\rangle$, $|1\rangle$, $|+\rangle$, $|-\rangle$
. Dealer and \rm{P}$_t$ then jointly detect the eavesdropping in the quantum channel.

Specifically, Dealer waits for \rm{P}$_t$ to received particles completely. After confirming the location of encrypted secret in received sequence with Dealer, \rm{P}$_t$ randomly measures the other particles in the basis $\{|0\rangle,|1\rangle\}$ or $\{|+\rangle,|-\rangle\}$. Then Dealer publishes the correct basis and \rm{P}$_t$ public the measurement. When fault measurements exceeding the number set by Dealer, Dealer determines the existence of eavesdropping. If there exists eavesdropping, the protocol will be  terminated. Conversely, the protocol  continues.

\subsection{Secret Reconstruction Phase}

\textbf{(R$_1$)} Other user \rm{P}$_u$ chooses an arbitrary angle $\delta _u^j$ for each quantum secret $|\psi {\rangle _j}$, ensuring that the angle is known only to himself, where $u \in \{ 1,2,...,t - 1\} $.

\textbf{(R$_2$)} The secret reconstructor \rm{P}$_t$ prepares $m(t-1)$ two-particle cluster states $|T\rangle^j_u = CZ|+\rangle|+\rangle$, which are utilized to teleport information for decrypting $m$ encrypted quantum secrets $|\Psi\rangle_j$. For each $|T\rangle^j_u$,  reconstructor \rm{P}$_t$ transmits one particle of  $|T\rangle^j_u$ with decoy particles to the user ${{\rm{P}}_u}$. After receiving, they jointly detect eavesdropping.

\textbf{(R$_3$)} The user ${{\rm{P}}_u}$ measures particles sent by \rm{P}$_t$ in the basis $\{ |{0_{ - \delta _u^j}}\rangle ,|{1_{ - \delta _u^j}}\rangle \} $ and announces the measurement result ${m_{\delta _u^j}} \in \{ 0,1\} $. The secret reconstructor  \rm{P}$_t$ then performs the operation ${Z^{{m_{\delta _u^j}}}}$ on the remaining particle of two-particle cluster states $|T\rangle^j_u$ and obtain the quantum state $|{ + _{\delta _u^j}}\rangle $, denoted as $|\Delta \rangle _u^j $, where
$u \in \{ 1,2,...,t - 1\} $ and $j \in \{ 1,2,...,m\} $.

\begin{figure}[h]
    \centering
    \includegraphics[width=1 in]{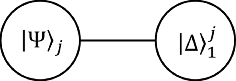}
    \caption{The secret reconstructor  \rm{P}$_t$ entangles $|\Psi {\rangle _j}$ and $|\Delta \rangle _1^j$ into two-particle cluster states, where the lines between particles represent the operation $CZ$.}
    \label{fig2}
\end{figure}

\textbf{(R$_4$)} The secret reconstructor \rm{P}$_t$ entangles $|\Psi {\rangle _j}$ and $|\Delta \rangle _1^j$ into two-particle cluster state $|C\rangle_{D,1}^j$ by performing the operation $CZ_{D,1}$, as shown in Figure \ref{fig2}. Please note that the cluster state $|C\rangle$ here is different from the former cluster state $|T\rangle$ used for teleportation.

\textbf{(R$_5$)} The secret reconstructor \rm{P}$_t$ measures the first particle of the cluster state in Figure \ref{fig2} (\ie, the particle denoted as $|\Psi {\rangle _j}$) in the basis $\{ | + \rangle ,| - \rangle \} $. When the state is $| + \rangle$, the measurement result is noted as $m_1^j = 0$.  When the state is $| - \rangle$, the result is noted as $m_1^j = 1$. Then \rm{P}$_t$ publishes the measurement result $m_1^j$ to the user \rm{P}$_1$. \rm{P}$_1$ calculates the rotation angle $\sigma _1^j = {( - 1)^{m_1^j + 1}}\delta _1^j + \gamma _1^j$ based on $m_1^j$ and publishes it to reconstructor \rm{P}$_t$. \rm{P}$_t$ then performs the operation
${R_X}(\sigma _1^j)H{X^{m_1^j}}$ on the second particle of cluster state. Now the second particle is denoted as $|\Delta '\rangle _1^j$.

\begin{figure}[h]
    \centering
    \includegraphics[width=1 in]{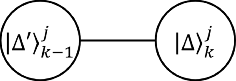}
    \caption{The secret reconstructor \rm{P}$_t$ entangles $|\Delta '\rangle _{k-1}^j$ and $|\Delta \rangle _k^j$ into two-particle cluster states, where the lines between particles represent the operation $CZ$.}
    \label{fig3}
\end{figure}

\textbf{(R$_{k+4}$, $k=2,3...,t-1$)} The secret reconstructor \rm{P}$_t$ entangles $|\Delta '\rangle _{k-1}^j$ and $|\Delta \rangle _k^j$ into cluster state $|C\rangle_{k-1,k}^j$ by performing the operation $CZ_{k-1,k}$, as shown in Figure \ref{fig3}.
Then \rm{P}$_t$ measures the first particle of the cluster state in Figure \ref{fig3} (\ie, the particle denoted as $|\Delta '\rangle _{k-1}^j$) in the basis $\{ | + \rangle ,| - \rangle \} $. \rm{P}$_t$ publishes the measurement result $m_k^j$ to the user \rm{P}$_k$. \rm{P}$_k$ calculates the rotation angle $\sigma _k^j = {( - 1)^{m_k^j + 1}}\delta _k^j + \gamma _k^j$ based on $m_k^j$ and publishes it to \rm{P}$_t$. \rm{P}$_t$ then performs the operation
${R_X}(\sigma _k^j)H{X^{m_k^j}}$ on the second particle of cluster state. And the second particle is now denoted as $|\Delta '\rangle _k^j$.

\textbf{(R$_{t+4}$)} Now, the secret reconstructor \rm{P}$_t$ can obtain the quantum state
\begin{equation}
    \begin{aligned}
            |\Psi '{\rangle _j} = {R_X}(\mathop \sum \limits_{k = 1}^{t - 1} \gamma _k^j)|\Psi {\rangle _j}
         = {R_X}(\mathop \sum \limits_{k = 1}^{t - 1} \gamma _k^j + \gamma _D^j)|\psi {\rangle _j}.
    \end{aligned}
\end{equation}
Then, \rm{P}$_t$ performs the operation $R_X(\gamma^j_t)$ on the $|\Psi'\rangle_j$, he can obtain the quantum state
\begin{equation}
    \begin{aligned}
        |\Psi ''{\rangle _j} = {R_X}(\mathop \sum \limits_{k = 1}^{t} \gamma _k^j + \gamma _D^j)|\psi {\rangle _j}=|\psi\rangle_j,
    \end{aligned}
\end{equation}
where the operation $R_X(\mathop \sum \limits_{k = 1}^{t} \gamma _k^j + \gamma _D^j)$  according to Equation (\ref{sumeq2pi}) is equal to $I$ (ignore the global phase).

\subsection{A simple example}\label{example}
As an example of (3,4) threshold quantum multi-secret sharing, the protocol has four users Alice, Bob, Charlie and Felix in addition to Dealer, where any three of users can cooperate to obtain multiple quantum secrets. Suppose that $\rm{q}=7$, $x_A=1$, $x_B=3$, $x_C=5$, and $x_F=6$. The number chosen by Dealer is $s_D=4$, then the polynomial is $f(x) = 7 - 4 + 2x + {x^2} = 3 + 2x + {x^2}$. Alice's share is ${s_A} = f(x_A)\mod 7 = 6$, Bob's share is ${s_B} = f(x_B)\mod 7 = 4$, Charlie's share is  ${s_C}\  = f(x_C)\mod 7 = 3$, and Felix's share is  ${s_F} = f(x_F)\mod 7 = 2$.

Without loss of generality, suppose that  Alice, Bob and Charlie cooperate to reconstruct the secret quantum sequence $\{|\psi\rangle_1,|\psi\rangle_2\}$, where Charlie is the secret reconstructor and the quantum secrets are $|\psi {\rangle _1} = 1/2|0\rangle  + \sqrt 3 /2|1\rangle  $ and $|\psi {\rangle _2} = |0\rangle $. First, Dealer announces ${w_1} = 1$ and ${w_2} = 2$.
Alice, Bob and Charlie calculate their parameters according to Equation (\ref{ci}) as  ${c_A} = 6$,  ${c_B} = 2$ and  ${c_C} = 2$, respectively. Then Dealer and these three users calculate the angles according to Equation (\ref{angle}) as $\gamma _D^1 = 8\pi /7$,  $\gamma _A^1 = 12\pi /7$, $\gamma _B^1 = 4\pi /7$ ,$\gamma _C^1 = 4\pi /7$ and $\gamma _D^2 = 16\pi /7$,  $\gamma _A^2 = 24\pi /7$, $\gamma _B^2 = 8\pi /7$ ,$\gamma _C^2 = 8\pi /7$. Dealer encrypts quantum secrets ${\{ |\psi \rangle _1},|\psi {\rangle _2}\} $ as ${\{ |\Psi \rangle _1=R_X(\gamma _D^1)|\psi\rangle_1},|\Psi {\rangle _2}=R_X(\gamma _D^2)|\psi\rangle_2\} $, and sends them to the secret reconstructor Charlie.

Alice and Bob randomly choose two angles  $\delta _A^1 = 2\pi /7$ and  $\delta _A^2 = \pi /7$,  $\delta _B^1 = 4\pi /7$ and  $\delta _B^2 = 3\pi /7$, respectively. To decrypt encrypted quantum secrets $\{|\Psi\rangle_1,|\Psi\rangle_2\}$, Charlie prepares four two-particle cluster states. These four cluster states are noted as $|T\rangle^1_A, |T\rangle^2_A, |T\rangle^1_B$ and $|T\rangle^2_B$. Charlie sends one particle of $|T\rangle^1_A$ and one of $|T\rangle^2_A$ to Alice, and sends one particle of $|T\rangle^1_B$ and one of $|T\rangle^2_B$ to Bob.  Alice measures the particles in the basis $\{ |{0_{ - 2\pi /7}}\rangle ,|{1_{ - 2\pi /7}}\rangle \} $ and basis $\{ |{0_{ - \pi /7}}\rangle ,|{1_{ - \pi /7}}\rangle \} $, respectively. Bob measures the particles in the basis $\{ |{0_{ - 4\pi /7}}\rangle ,|{1_{ - 4\pi /7}}\rangle \} $ and basis $\{ |{0_{ - 3\pi /7}}\rangle ,|{1_{ - 3\pi /7}}\rangle \} $, respectively. They publish the corresponding measurement results ${m_{\delta _k^j}}$, where $j \in \{ 1,2\} $ and $k \in \{ A,B\} $. Charlie performs operations ${Z^{{m_{\delta _k^j}}}}$ based on the measurement results ${m_{\delta _k^j}}$ to obtain the particles $|\Delta \rangle _A^1 = |{ + _{2\pi /7}}\rangle $, $|\Delta \rangle _A^2 = |{ + _{\pi /7}}\rangle$, $|\Delta \rangle _B^1 = |{ + _{4\pi /7}}\rangle $and $|\Delta \rangle _B^2 = |{ + _{3\pi /7}}\rangle $.

To reconstruct the quantum secret $|\psi {\rangle _1}$, Charlie performs the operation $CZ^1_{D,A}$ to entangle $|\Psi {\rangle _1}$ and $|\Delta \rangle _A^1$ into cluster state $|C\rangle _{D,A}^1$.
Next, Charlie measures the first particle of cluster state $|C\rangle _{D,A}^1$ in the basis $\{ | + \rangle ,| - \rangle \} $ and announces the measurement. If the measurement result $m_A^1 = 0$, Alice computes the angle $\sigma _A^1 =  - 2\pi /7 + 12\pi /7 = 10\pi /7$ and publishes it to Charlie, who operates ${R_X}(10\pi /7)H$ on the second particle of $|C\rangle _{D,A}^1$.
If the measurement result $m_A^1 = 1$, Alice computes the angle $\sigma _A^1 = 2\pi /7 + 12\pi /7 = 2\pi $ and publishes it to Charlie, who operates ${R_X}(2\pi )HX$ on the second particle of $|C\rangle _{D,A}^1$. The second particle is denoted as $|\Delta '\rangle _A^1$.

Charlie performs the operation $CZ^1_{A,B}$ to entangle the particles $|\Delta '\rangle _A^1$ and $|\Delta \rangle _B^1$ into cluster states $|C\rangle _{A,B}^1$. Then, Charlie measures the first particle of $|C\rangle _{A,B}^1$ in the basis $\{ | + \rangle ,| - \rangle \} $ and announces the measurement result. If the measurement result $m_B^1 = 0$, Bob computes the angle $\sigma _B^1 =  - 4\pi /7 + 4\pi /7 = 0$ and publishes it to Charlie, who operates ${R_X}(0)H$ on the second particle of $|C\rangle _{A,B}^1$. If the measurement result is $m_B^1 = 1$, Bob computes the result $\sigma _B^1 = 4\pi /7 + 4\pi /7 = 8\pi /7$ and publishes it to Charlie, who operates ${R_X}(8\pi /7)HX$ on the second particle of $|C\rangle _{A,B}^1$, obtaining the particle $|\Psi '{\rangle _1}$. Finally, Charlie performs the  operation  ${R_X}(\gamma _C^1) = {R_X}(4\pi /7)$ on the particle $|\Psi '{\rangle _1}$ to recover the original quantum secret $|\psi {\rangle _1} = 1/2|0\rangle  + \sqrt 3 /2|1\rangle $ (ignore the global phase).

The reconstruction of quantum secret $|\psi {\rangle _2}$ is similar. The secret reconstructor Charlie performs the operation $CZ^2_{D,A}$ to entangle $|\Psi {\rangle _2}$ and $|\Delta \rangle _A^2$ into cluster states $|C\rangle _{D,A}^2$. Then, Charlie measures the first particle of $|C\rangle _{D,A}^2$ in the basis $\{ | + \rangle ,| - \rangle \} $ and announces the measurement result. If $m_A^2 = 0$, Alice computes the angle $\sigma _A^2 =  - \pi /7 + 24\pi /7 = 23\pi /7$ and publishes it to Charlie, Charlie performs the operation ${R_X}(23\pi /7)H$ on the second particle of $|C\rangle _{D,A}^2$.
If $m_A^2 = 1$, Alice computes the angle and publishes it to Charlie, Charlie performs the operation ${R_X}(25\pi /7)HX$ on the second particle of $|C\rangle _{D,A}^2$, obtaining the particle $|\Delta '\rangle _A^2$. Next, Charlie entangles $|\Delta '\rangle _A^2$ and $|\Delta \rangle _B^2$ into cluster state $|C\rangle _{A,B}^2$, and measures the first particle of $|C\rangle _{A,B}^2$ in the basis $\{ | + \rangle ,| - \rangle \} $. Then Charlie announces the measurement result $m_B^2$.
If $m_B^2 = 0$, Bob computes the angle $\sigma _B^1 =  - 3\pi /7 + 8\pi /7 = 5\pi /7$ and publishes it to Charlie, who operates ${R_X}(5\pi /7)H$ on the second particle of $|C\rangle _{A,B}^2$.
If $m_B^2 = 1$, Bob computes the angle $\sigma _B^1 = 3\pi /7 + 8\pi /7 = 11\pi /7$ and publishes it to Charlie, who operates ${R_X}(11\pi /7)HX$ on the second particle of $|C\rangle _{A,B}^2$,  obtaining $|\Psi '{\rangle _2}$. Finally, Charlie performs the operation ${R_X}(\gamma _C^2)={R_X}(8\pi /7)$ on the particle $|\Psi '{\rangle _2}$ to recover the original quantum secret $|\psi {\rangle _2} = |0\rangle $ (ignore the global phase). For clarity, the evolution of quantum states of the system is shown in the Table \ref{tab}.

\begin{table*}[!t]
    \caption{Concrete process of the example\label{tab}}
    \centering
        \begin{tabular}{c|c|c}
            \hline
            Concrete process of the example & Quantum state of system 1 & Quantum state of system 2 \\
            \hline
            \makecell[c]{Alice, Bob and Charlie cooperate\\ to reconstruct quantum secrets, \\Dealer transmit encrypted \\quantum secrets $\{|\Psi\rangle_1,|\Psi\rangle_2\}$ \\to secret reconstructor Charlie} & \makecell[c]{$|\Psi\rangle_1=R_X(8\pi/7)|\psi\rangle_1$} & \makecell[c]{$|\Psi\rangle_2=R_X(16\pi/7)|\psi\rangle_2$}\\
            \hline
            \makecell[c]{Charlie entangles four two-particle \\cluster states and sends one \\particle of each of them to \\Alice and Bob, respectively.} &\makecell[c]{$|\Psi\rangle_1\otimes|T\rangle^1_A\otimes|T\rangle^1_B$} & \makecell[c]{$|\Psi\rangle_2\otimes|T\rangle^2_A\otimes|T\rangle^2_B$}\\
            \hline
            \makecell[c]{Alice measures the particles  sent by\\ Charlie in the basis $\{ |{0_{ - 2\pi /7}}\rangle ,|{1_{ - 2\pi /7}}\rangle \} $ \\ and $\{ |{0_{ - \pi /7}}\rangle ,|{1_{ - \pi /7}}\rangle \} $, respectively. \\Then Alice publishes the\\ measurement results ${m_{\delta _A^1}}$ and ${m_{\delta _A^2}}$.\\ Charlie performs operations ${Z^{{m_{\delta _A^1}}}}$\\ and ${Z^{{m_{\delta _A^2}}}}$ to obtain the particles\\ $|\Delta \rangle _A^1 = |{ + _{2\pi /7}}\rangle $ and $|\Delta \rangle _A^2 = |{ + _{\pi /7}}\rangle$.} & \makecell[c]{$|\Psi\rangle_1\otimes|T \rangle _A^1\otimes|T\rangle^1_B$\\ $\;\downarrow$ Alice measures\\$|\Psi\rangle_1\otimes|\Delta \rangle _A^1\otimes|T\rangle^1_B$
            } &
            \makecell[c]{$|\Psi\rangle_2\otimes|T \rangle _A^2\otimes|T\rangle^2_B$\\ $\;\downarrow$ Alice measures\\$|\Psi\rangle_2\otimes|\Delta \rangle _A^2\otimes|T\rangle^2_B$
             }\\
            \hline
            \makecell[c]{Bob measures the particles sent by \\ Charlie  in the basis $\{ |{0_{ - 4\pi /7}}\rangle ,|{1_{ - 4\pi /7}}\rangle \} $\\ and  $\{ |{0_{ - 3\pi /7}}\rangle ,|{1_{ - 3\pi /7}}\rangle \} $,\\ respectively. Then Bob publishes the\\ measurement results ${m_{\delta _B^1}}$ and ${m_{\delta _B^2}}$.\\ Charlie performs operations ${Z^{{m_{\delta _B^1}}}}$\\ and ${Z^{{m_{\delta _B^2}}}}$ to obtain the particles\\ $|\Delta \rangle _B^1 = |{ + _{4\pi /7}}\rangle $ and $|\Delta \rangle _B^2 = |{ + _{3\pi /7}}\rangle $.} & \makecell[c]{
                $\downarrow$ Bob measures\\$|\Psi\rangle_1\otimes|\Delta\rangle^1_A\otimes|\Delta \rangle _B^1$\\
                } &
            \makecell[c]{
                $\downarrow$ Bob measures\\$|\Psi\rangle_2\otimes|\Delta \rangle _A^2\otimes|\Delta\rangle^2_B$
                \\
                }\\
            \hline
            \makecell[c]{Charlie performs the operations \\$CZ^j_{D,A}$ to entangle $|\Psi {\rangle _j}$ and $|\Delta \rangle _A^j$\\ into cluster state $|C\rangle _{D,A}^j$, where $j\in\{1,2 \}$. \\ Charlie measures the first particle of $|C\rangle _{D,A}^j$\\in the basis $\{ | + \rangle ,| - \rangle \} $. Alice computes \\the angle $\sigma _A^j = {( - 1)^{m_A^j + 1}}\delta _A^j + \gamma _A^j $ \\according to the measurement result $m_A^j$.\\ Charlie operates ${R_X}(\sigma _A^j)H{X^{m_A^j}}$ on the\\ second particle   obtaining the particle $|\Delta '\rangle _A^j$.}
            & \makecell[c]{
                $\downarrow$ Charlie entangles\\
                $|C\rangle _{D,A}^1\otimes|\Delta\rangle^1_B$
                \\$\downarrow$ then measures and operates
                \\$|\Delta{'}\rangle_{A}^1\otimes|\Delta\rangle^1_B$
                }&\makecell[c]{$\downarrow$ Charlie entangles\\
                $|C\rangle _{D,A}^2\otimes|\Delta\rangle^2_B$
                \\$\downarrow$ then measures and operates
                \\$|\Delta{'}\rangle_{A}^2\otimes|\Delta\rangle^2_B$
                }\\
            \hline
            \makecell[c]{
            Charlie performs the operations \\$CZ^j_{A,B}$ to entangle $|\Delta '\rangle _A^j$ and $|\Delta \rangle _B^j$\\ into cluster state $|C\rangle _{A,B}^j$, where $j\in\{1,2 \}$. \\ Charlie measures the first particle of $|C\rangle _{A,B}^j$\\in the basis $\{ | + \rangle ,| - \rangle \} $. Bob computes \\the angle $\sigma _B^j = {( - 1)^{m_B^j + 1}}\delta _B^j + \gamma _B^j $ \\according to the measurement result $m_B^j$.\\ Charlie operates ${R_X}(\sigma _B^j)H{X^{m_B^j}}$ on the\\ second particle obtaining the particle $|\Psi'\rangle_j$.}
            & \makecell[c]{
                $\downarrow$ Charlie entangles\\
                $|C\rangle _{A,B}^1$
                \\$\downarrow$ then measures and operates
                \\$|\Psi{'}\rangle_1$
                }&\makecell[c]{$\downarrow$ Charlie entangles\\
                $|C\rangle _{A,B}^2$
                \\$\downarrow$ then measures and operates
                \\$|\Psi{'}\rangle_2$
                }\\
            \hline
            \makecell[c]{Charlie performs the operations \\$R_X(\gamma^1_C)$ and $R_X(\gamma^2_C)$ on $|\Psi'\rangle_1$ \\and $|\Psi'\rangle_1$to recover the original \\quantum secrets $|\psi\rangle_1$ and $|\psi\rangle_2$, \\ignoring the global phase.}
            & \makecell[c]{$\downarrow$ Charlie operates\\${R_X}(4\pi)|\psi {\rangle _1}=|\psi\rangle_1$} & \makecell[c]{$\downarrow$ Charlie operates\\${R_X}(8\pi)|\psi {\rangle _2}=|\psi\rangle_2$}\\
            \hline
        \end{tabular}
\end{table*}

\section{Performance analysis}\label{Sec4}

In this section, the correctness and security analysis of the proposed protocol is given. In the security analysis, the internal and external attacks is discussed to prove the security of the protocol.

\subsection{Correctness}
Firstly, in order to decrypt the encrypted quantum secret $|\Psi {\rangle _j}$, the secret reconstructor \rm{P}$_t$ prepares $m\cdot(t-1)$ two-particle cluster states. Then \rm{P}$_t$ sends one particle of each of them to other user \rm{P}$_u$, who measures the particle in the basis $\{ |{0_{ - {\delta ^j_u}}}\rangle ,|{1_{ - {\delta ^j_u}}}\rangle \} $, where $u\in\{1,2,\cdots,t-1\}$. These two-particle cluster states satisfies
\begin{equation}
    \begin{aligned}
        |T\rangle_u^j=C{Z}| + {\rangle }| + {\rangle} = \frac{1}{\sqrt{2}}(|{0_{ - \delta _u^j}}\rangle |{ + _{\delta _u^j}}\rangle  + |{1_{ - \delta _u^j}}\rangle |{ - _{\delta _u^j}}\rangle).
    \end{aligned}
    \label{measurein01basis}
\end{equation}
According to Equation (\ref{measurein01basis}), the quantum state $|{ + _{\delta _u^j}}\rangle$ or $|{ -_{\delta _u^j}}\rangle$ can be obtained after the user measuring any one particle of $|T\rangle_u^j$. The user \rm{P}$_u$ publishes the measurement result ${m_{\delta _u^j}} \in \{ 0,1\} $. If the obtained quantum state after measuring is $|{0_{ - \delta _u^j}}\rangle$, the measurement result is noted as ${m_{\delta _u^j}}=0$. If the obtained quantum state after measuring is $|{1_{ - \delta _u^j}}\rangle$, the measurement result is ${m_{\delta _u^j}}=1$. The secret reconstructor \rm{P}$_t$ then performs quantum operations ${Z^{{m_{\delta _u^j}}}}$ on the remaining particle of the two-particle cluster state.
Hence the secret reconstructor will obtain the quantum state $|{ + _{\delta _u^j}}\rangle$, denoted as $|\Delta\rangle^j_u=|{ + _{\delta _u^j}}\rangle$. Subsequently, the secret reconstructor \rm{P}$_t$ entangles $|\Psi {\rangle _j}$ and $|\Delta \rangle _1^j$ into the cluster states $|C{\rangle _{D,1}}$ by performing the operation $CZ_{D,1}$, where the subscripts $D$ and $1$ indicate that the particles come from Dealer and the user \rm{P}$_1$. The two-particle cluster state $|C{\rangle _{D,1}}$ satisfies
\begin{equation}
    \begin{aligned}
        |C{\rangle _{D,1\;\;\;}} =& C{Z_{D,1}}|\Psi {\rangle _j}|\Delta \rangle _1^j\\
             =& C{Z_{D,1}}|\Psi {\rangle _j}|{ + _{\delta _1^j}}\rangle \\
             = & \frac{1}{{\sqrt 2 }}| + \rangle  \otimes {({R_Z}(\delta _1^j)H|\Psi \rangle _j})+ \\
            & \frac{1}{{\sqrt 2 }}| - \rangle  \otimes {({R_Z}(\delta _1^j)XH|\Psi \rangle _j}).
    \end{aligned}
\end{equation}

After measuring the first particle of two-particle cluster state $|C{\rangle^j_{D,1}}$ in the basis $\{ | + \rangle ,| - \rangle \} $, the second particle will be collapsed to the quantum state ${R_Z}(\delta _1^j){X^{m_1^j}}H|\Psi {\rangle _j}$, where $m_1^j$ represents the measurement result, which is published by the secret reconstructor \rm{P}$_t$. Then the user \rm{P}$_1$ calculates $\sigma _1^j = {( - 1)^{m_1^j + 1}}\delta _1^j + {\gamma _1}$ and publishes it to the secret reconstructor \rm{P}$_t$.
The secret reconstructor \rm{P}$_t$ performes the operation ${R_X}(\sigma _1^j)H{X^{m_1^j}} = {R_X}({( - 1)^{m_1^j + 1}}\delta _1^j + {\gamma _1})H{X^{m_1^j}}$ on the second particle, the state of the second particle will be changed to ${R_X}({( - 1)^{m_1^j + 1}}\delta _1^j + {\gamma _1})H{X^{m_1^j}}{R_Z}(\delta _1^j){X^{m_1^j}}H|{\Psi _j}\rangle$. According to Equation (\ref{eqcite1}), the quantum state can be simplified to ${R_X}({\gamma _1})|{\Psi _j}\rangle$ that is noted as $|\Delta'\rangle^j_1$.
Next, the secret reconstructor \rm{P}$_t$ entangles $|\Delta'\rangle^j_{k-1}$ and $|\Delta\rangle^j_k$ into the cluster states  $|C\rangle_{k-1,k}^j$ by performing the operation $CZ_{k-1,k}$. Here, number $k$ ranges over the set $\{2,3,...,t-1\}$. The secret reconstructor \rm{P}$_t$ measures the first particle of $|C\rangle_{k-1,k}^j$ in the basis $\{ | + \rangle ,| - \rangle \} $ and publishes the measurement results ${m_k^j}$ to the user \rm{P}$_k$.
After obtaining the angle $\sigma _k^j$, \rm{P}$_t$ performs operations ${R_X}(\sigma _k^j)H{X^{m_k^j}}$ on the second particle obtaining the quantum state ${R_X}({\gamma _k})|\Delta '\rangle _{k - 1}^j$. After the above cycle, eventually the secret reconstructor \rm{P}$_t$ obtains the quantum state
\begin{equation}
    \begin{aligned}
        |\Psi '{\rangle _j} = {R_X}(\mathop \sum \limits_{k = 1}^{t - 1} {\gamma _k})|\Psi {\rangle _j}
        = {R_X}({\gamma _D} + \mathop \sum \limits_{k = 1}^{t - 1} {\gamma _k})|\psi {\rangle _j}.
    \end{aligned}
\end{equation}
Next, the reconstructor applies a rotation operation $R_X(\gamma^j_C)$ to the particle $|\Psi '{\rangle _j}$ obtaining the quantum state
\begin{equation}
    \begin{aligned}
        |\Psi ''{\rangle _j} ={R_X}({\gamma _t})|\Psi'\rangle _j
        = {R_X}({\gamma _D} + \mathop \sum \limits_{k = 1}^t {\gamma _k})|\psi {\rangle _j}.
    \end{aligned}
\end{equation}
According to Equation (\ref{sumeq2pi}), the above state is equal to the original quantum secret (ignore the global phase). In summery, the secret reconstructor \rm{P}$_t$ can reconstruct correct quantum secrets $|\psi {\rangle _j}$ and the protocol satisfies the correctness.

Besides, we note that the correctness proof of Ref.\cite{MaRH2024} introduces the conclusion of Nielsen \cite{Nielsen2006} that the measurement can be delayed until after full entanglement in the cluster states. However, there are some additional quantum operations, such as ${R_X}$, $H$ and ${X}$, etc., are performed in Ref.\cite{MaRH2024}. These operations cannot commute with the operation $CZ$. Thus the delay of the measurement until after all particles have been entangled will not satisfy correctness. But it is worth noting that the quantum experiments in Ref.\cite{MaRH2024} still satisfy correctness, because the entanglements and measurements are performed in form of two-particle cluster state like this paper. Therefore, the correctness can still be satisfied by simply adapting the secret reconstruction process of protocol\cite{MaRH2024} in the two-particle cluster state.

\subsection{Security}
In this section, several common attack  will be discussed, comprising the external attacks and internal attacks. And the attack targets will be categorized as attacks on Dealer, attacks on the secret reconstructor and attacks on other users. This section will prove that none of these attacks can steal secrets, proving that this protocol is secure against several common external attacks and internal attacks.

\subsubsection{External Attacks}

Suppose there is an external eavesdropper Eve who attempts to eavesdrop on the quantum secrets. She can first choose to intercept the particles sent by Dealer, \ie, $|\Psi {\rangle _j}$, where $j \in \{ 1,2,...,m\} $. But due to the decoy particles, Eve's eavesdropping will introduce errors in the detection. Assuming that the number of decoy particles in $X$ basis is $d_1$, and the number in $Z$ basis is $d_2$.
When intercepting the secret, Eve does not know the measurement basis of the decoy particles. Therefore, she can only randomly choose one of the measurement basis of $X$ or $Z$ basis to measure. After the measurement, Eve then prepares fake particles to replace the original particles according to the measurement results.
It can be calculated that the probability that Eve does not introduce any error after sending the fake particles is \\${(1/2 \times 1 + 1/2 \times 1/2)^{{d_1}+{d_2}}} = {(3/4)^{{d_1}+{d_2}}}$.
When the number of decoy particles is larger than a certain number, the probability that Eve does not introduce any error will tend to $0$. Once the eavesdropping is detected, the protocol will be terminated. Even if Eve introduces errors regardless, she will only have access to the encrypted quantum secrets $|\Psi {\rangle _j}$, instead of the original quantum secret $|\psi {\rangle _j}$ sent by Dealer. Therefore, she cannot steal any useful information of the quantum secrets.

If Eve attempts to steal the share $s_k$ of user \rm{P}$_k$, where $k \in\{ 1,2,...,t-1\}$. Then she can either intercept the particles sent to the user by the secret reconstructor \rm{P}$ _t$, or steal the information in the public classical channel. On the one hand, Eve intercepts particles in the quantum channel, her eavesdropping will be detected due to the decoy particles, leading to the termination of the protocol. Moreover, she will only have access to some unmeasured cluster particles with no valid information. On the other hand, Eve steals on the public channel, there are two kinds of information that she can steal, measurement results and angle information. However, neither of these two types of information can reveal the user's share $s_k$. Hence, Eve cannot steal the user's share.

Finally, consider the case where Eve attempts to steal the share of secret reconstructor \rm{P}$_t$. Since the reconstructor does not reveal any information about his share, Eve cannot steal it. In conclusion, the external eavesdropper  cannot steal any useful information in the presented protocol. The protocol is secure against external attacks.

\subsubsection{Internal Attacks}
In this subsection, we will analyze three types of internal attacks, the reconstructor attacks, the other user attacks and the collusion attacks. Regarding collusion attacks, we will consider two extreme cases where only one participant is honest, only the reconstructor \rm{P}$_t$ is honest and only one non-reconstructor user is honest.

\paragraph{Reconstructor Attacks}
Since the secret reconstructor \rm{P}$_t$ is the participant of protocol, his eavesdropping will not be detected by the decoy particle detection. Therefore, the reconstructor attacks are more harmful to the protocol. For a dishonest reconstructor, his aim is to steal the shares $s_u$ of other users \rm{P}$_u$, where $u\in\{1,2,...,t -1\}$. Then he can choose to steal the secret in the following way. Since the other users \rm{P}$_u$ have to publish different angles $\sigma^j_u$ for different secrets $|\psi\rangle_j$ based on the reconstructor's measurement results $m_u^j$, the dishonest reconstructor \rm{P}$_t^{'}$ can publishe fake measurement results, such as ${m'}_u^j\in\{0,1\}$, to induce \rm{P}$_u$ publishing the angles $ {\sigma'}_u^j = \gamma _u^j + {( - 1)^{{m'}_u^j}}\delta _u^j$. Then the reconstructor can operate ${R_X}({\sigma'}_u^j)$ on $|\Delta \rangle _u^j$ to obtain the particles $|\Delta ''\rangle _u^j = {R_Z}(\gamma _u^j)| + \rangle$. In this form, other angles are stripped from the related angles $\gamma _u^j$ of shares.
However, since the reconstructor does not know the correct measurement basis of $|\Delta ''\rangle _u^j = {R_Z}(\gamma _u^j)| + \rangle$, and the angle of each quantum secret is encrypted by different parameters $\omega_j$, the dishonest reconstructor cannot obtain any valid information by this eavesdropping approach. In addition, even if the dishonest reconstructor \rm{P}$_t$ steals the secret based on the angles $ {\sigma'}_u^j = \gamma _u^j + {( - 1)^{{m'}_u^j}}\delta _u^j$ announced in the public classical channel, he still cannot obtain any valid information because he does not know the angles $\delta_u^j$. In conclusion, the protocol is secure against internal attacks by the dishonest reconstructor.

\paragraph{Other User Attacks}
Without loss of generality, assume that the user \rm{P}$_1$ is a dishonest user,  who attempts to steal Dealer's quantum secrets, the shares of other users \rm{P}$_k$, and the share of the secret reconstructor \rm{P}$_t$, where $k=\{2,3,...,t-1\}$. First, the dishonest user \rm{P}$_1^{'}$ can intercept the particles sent by Dealer to steal Dealer's quantum secrets. But due to the presence of decoy particles, the intercepting will introduce errors that leads to the termination of the protocol. Even if the dishonest user \rm{P}$_1^{'}$ insists to steal the corresponding encrypted quantum secrets, he cannot decrypt them. Second, consider the case where the dishonest user \rm{P}$_1^{'}$ attacks on other user's share. If \rm{P}$_1^{'}$ intercepts the particles transmitted by the secret reconstructor \rm{P}$_t$, his attack will be detected and cause the protocol to terminate due to the decoy particles. He will only steal the unmeasured cluster particles and will obtain no useful information. If the dishonest user \rm{P}$_1^{'}$ eavesdrops on the public information, similar to external attacks, his attack will not obtain the shares of other users \rm{P}$_k$ since public information does not reveal the secret shares $s_k$. Finally, consider the attack on the secret reconstructor \rm{P}$_t$. Since the secrets reconstructor \rm{P}$_t$ does not reveal any information about his own shares in the protocol, the dishonest user \rm{P}$_1^{'}$ cannot steal the shares of the secret reconstructor \rm{P}$_t$. In conclusion, the protocol is secure against internal attacks by other user.

\paragraph{Collusion Attack}
A collusion attack refers to multiple participants working together to steal secrets.  The most extreme case of a collusion attack is where only one user is honest. If a protocol can withstand this most extreme attack, then it can withstand attacks with fewer collision attackers naturally. In this protocol, the  extreme collusion attacks can be classified into two categories, the collusion attack with the involvement of the secret reconstructor \rm{P}$_t$ and the collusion attack without the involvement of the secret reconstructor \rm{P}$_t$.

Let us first discuss the first type of collusion attack. Without loss of generality, assume that the only honest user is \rm{P}$_2$. All other users \rm{P}$_1$, \rm{P}$_3$, \rm{P}$_4$, $...$, \rm{P}$_{t-1}$ and the secret reconstructor \rm{P}$_t$ cooperate in the collusion attack. In this case, since all information is shared among the collusion attackers, the protocol is equivalent to a three-party agreement consisting of the honest Dealer, the honest user \rm{P}$_2$, and the dishonest reconstructor \rm{P}$_t^{'}$. Obviously, this situation is the internal attack which has been discussed. The dishonest secret recosntructor \rm{P}$_t^{'}$ cannot steal any useful information, and he cannot steal the quantum secrets.

The second type of collusion attack is that the reconstructor is the only one who is honest, and all other users \rm{P}$_1$, \rm{P}$_2$, $...$, \rm{P}$_{t-1}$ cooperate in the collusion attack. Then, all collusion attackers have two targets for stealing secrets, one is the quantum secret $|\Psi\rangle_j$ of Dealer and the other is the share of the secret reconstructor \rm{P}$_t$. When all the other users try to steal the quantum secret, they can choose to intercept the particles. Due to the decoy particles, their eavesdropping will introduce  errors. The protocol, therefore, will be terminated. And the collusion attackers cannot obtain any useful information. If the collusion attackers  attack on the share of the secret reconstructor \rm{P}$_t$, the collusion attacks cannot steal the share of secret reconstructor \rm{P}$_t$ since the reconstructor \rm{P}$_t$ never publishes his share. To sum up, the collusion attack cannot steal any useful information. Based on the above analysis, the protocol is secure against collusion attacks.

\section{Experiment}\label{Sec5}

\begin{figure*}[h]
    \centering
    \includegraphics[width=1.5 in, angle=270]{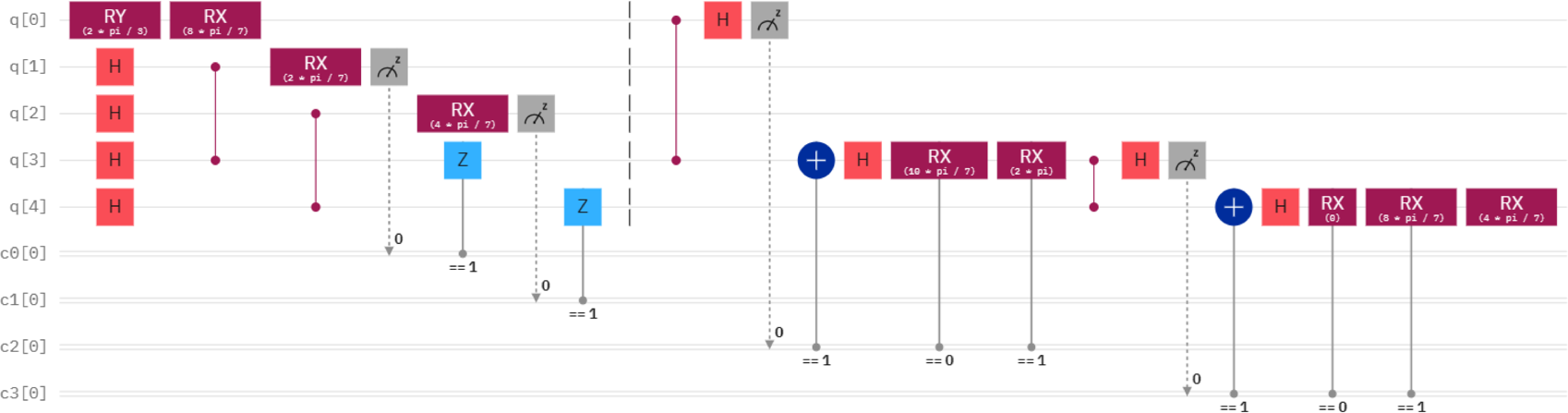}
    \caption{The circuit 1 that recovers quantum secret $|\psi\rangle_1$.}
    \label{circuit1}
\end{figure*}

To verify the correctness and feasibility of the protocol, a quantum simulation experiment on the quantum experimental platform IBM Q is conducted. The  experimental process are referred to the examples of the subsection \ref{example}. And the experimental circuit is shown in Figure \ref{circuit1}. On the circuit q[0], the quantum secret $|\psi {\rangle _1} = 1/2|0\rangle  + \sqrt 3 /2|1\rangle $ is generated by performing the quantum operation ${R_Y}(2\pi /3)$ since the initial state of each circuit is  $|0\rangle$.
Next, the operation ${R_X}(8\pi /7)$ is performed to encrypt the quantum secret $|\psi {\rangle _1}$ as $|\Psi {\rangle _1}$. On the circuits q[1], q[2], q[3] and q[4], the quantum states $|+\rangle$ are generated by four operations $H$. Then, the circuits q[1] and q[3], q[2] and q[4] are entangled into a two-particle cluster state, respectively. Followed by the operations ${R_X}(2\pi /7)$ and ${R_X}(4\pi /7)$ and the measurements in the basis $\{ |0\rangle ,|1\rangle \} $, which are equivalent to the measurements in the basis $\{ |{0_{ - 2\pi /7}}\rangle ,|{1_{ - 2\pi /7}}\rangle \} $ and the basis $\{ |{0_{ - 4\pi /7}}\rangle ,|{1_{ - 4\pi /7}}\rangle \} $. The measurement results are stored in the registers c0[0] and c1[0], respectively. Eventually, the operations $Z$ are performed based on the measurement results so that the quantum states $|{ + _{2\pi /7}}\rangle $ and $|{ + _{4\pi /7}}\rangle $ can be obtained in the circuits q[3] and q[4], respectively.

\begin{figure*}[h]
    \centering
    \includegraphics[width=1.5 in, angle=270]{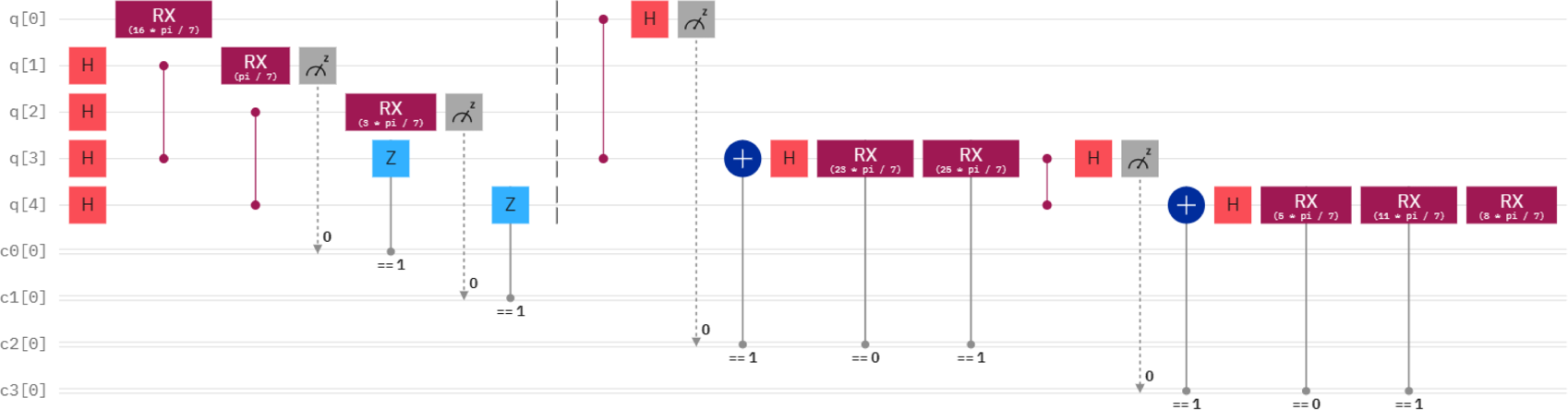}
    \caption{The circuit 2 that recovers quantum secret $|\psi\rangle_2$.}
    \label{circuit2}
\end{figure*}

Next, the circuits q[0] and q[3] are connected by operation $CZ_{D,A}$ to realise entanglement of $|\Psi {\rangle _1}$ and $|\Delta \rangle _A^1$ into cluster state $|C\rangle^1_{D,A}$. The operation $H$ is then performed on the circuit q[0] as well as the measurement in the basis $\{|0\rangle,|1\rangle\}$, which is equivalent to measurement in the basis $\{|+\rangle,|-\rangle\}$. The measurement result is stored in the register c2[0], and the quantum state $|\Delta '\rangle _A^1$ is obtained by the operation ${R_X}(10\pi /7)H$ or the operation ${R_X}(2\pi )HX$ based on the measurement result. The entanglement $CZ_{A,B}$ is performed on the circuits q[3] and q[4]. Then the measurement is performed in the basis $\{|+\rangle,|-\rangle\}$, and the operations ${R_X}(0)H$ or ${R_X}(8\pi /7)HX$ is applied on the circuit q[4] based on the measurement result. Finally, the quantum secret $|\psi\rangle_1$ can be recovered by operating ${R_X}({\gamma _C}) = {R_X}(4\pi /7)$ on the circuit q[4]. In addition, for the quantum secret $|\psi\rangle_2=|0\rangle$, the experimental circuit is similar as the one that recovers $|\psi\rangle_1$, as shown in the Figure \ref{circuit2}.

\begin{figure}[h]
    \centering
    \includegraphics[width=1.5 in, angle=270]{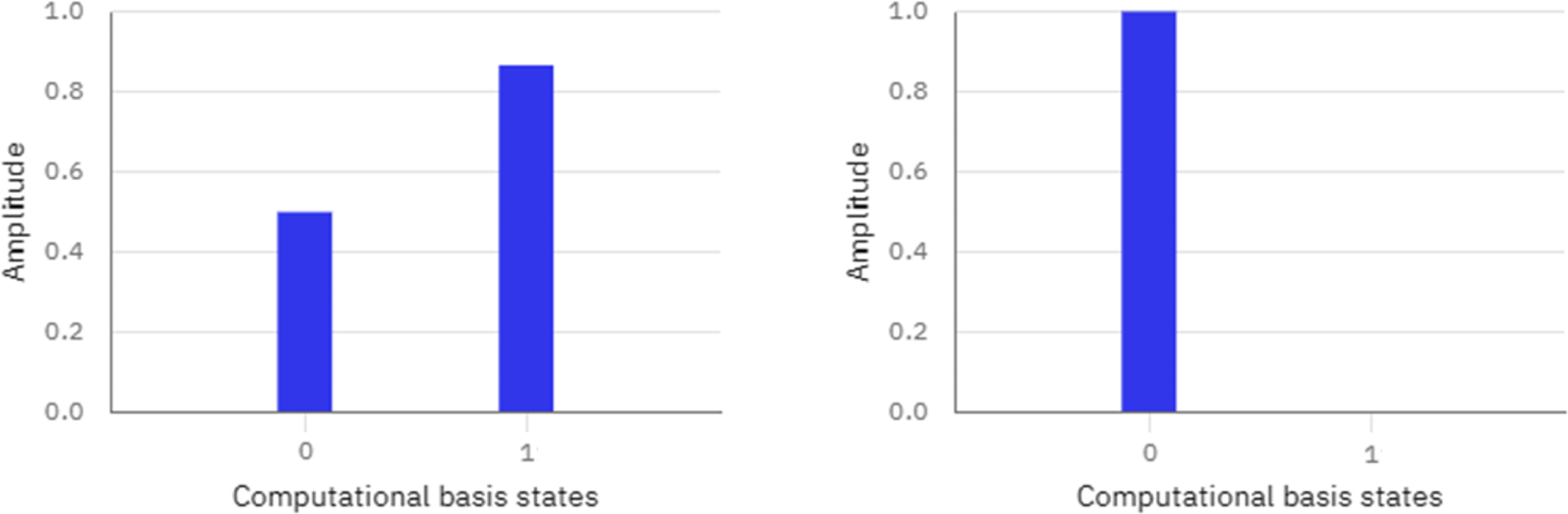}
    \caption{The measurement outcomes of the experiment.}
    \label{result}
\end{figure}

The experimental results of the above two circuits are shown in Figure \ref{result}. In Figure \ref{result}, the data below the horizontal axis represents the quantum state of q[4]. By observing the amplitudes of  Figure \ref{result}, it can be found that the quantum state recovered in left subgraph  is $1/2|0\rangle+\sqrt{3}/2|1\rangle$, which is equal to the original quantum secret $|\psi\rangle_1$. And the state in right subgraph   is $|0\rangle$, which is equal to the original quantum secret $|\psi\rangle_2$. In summary, the proposed protocol can correctly recover the original quantum secrets, proving that the protocol satisfies the correctness and feasibility.

\section{Conclusion}\label{Sec6}

A quantum ($t,n$) threshold multi-secret sharing protocol based on Lagrangian interpolation and cluster states is presented in this paper. Each share of the protocol is split into two parts, classical and quantum, for transmission, which protects the security of each quantum secret. By utilizing a new cluster state measurement basis, users except the secret reconstructor do not need to prepare quantum states, but only need to perform single particle measurements assisting secret reconstructor to recover quantum secrets. Meanwhile, the quantum operations used in this paper are  all common operations, and the secret dealer can be offline after transmitting the encrypted quantum secrets. Therefore, the proposed protocol is practical under current technological conditions. The security analysis shows that the proposed protocol is secure against several common external and internal attacks. Eventually, the quantum experiments on IBM Q platform prove that the presented protocol satisfies correctness and feasibility.

\section*{Acknowledgement}
This work was supported by National Natural Science Foundation of China (Grants No. 62171131 and No.61976053), Fujian Province Natural Science Foundation (Grants No. 2022J01186, No. 2023J01533 and No. 2022J05049), Fujian Province Department of Education Young and Middle aged Teacher Education Research Project (Grant No. JAT231018) and Program for New Century Excellent Talents in Fujian Province University.


\end{document}